\documentclass[
superscriptaddress,
 amsmath,amssymb,
 aps,
prl,
twocolumn,
]{revtex4-2}
\newcommand{\barium}{$^{133}$Ba$^+$}
\usepackage{braket}
\usepackage{graphicx}
\usepackage[colorlinks=true, allcolors=blue]{hyperref}
\usepackage{gensymb}
\usepackage{cancel}
\usepackage{upgreek}

\usepackage{float}  
\usepackage{xcolor}

\usepackage{soul}
\usepackage{braket}
\usepackage{graphicx}

\usepackage[colorlinks=true, allcolors=blue]{hyperref}

\usepackage{upgreek}

\begin{document}

\title{Eliminating qubit type cross-talk in the $omg$ protocol}
\author{Samuel R. Vizvary}
\email{samvizvary@g.ucla.edu}
\affiliation{Department of Physics and Astronomy, University of California Los Angeles, Los Angeles, CA, USA}
\author{Zachary J. Wall}
\affiliation{Department of Physics and Astronomy, University of California Los Angeles, Los Angeles, CA, USA}
\author{Matthew J. Boguslawski}
\affiliation{Department of Physics and Astronomy, University of California Los Angeles, Los Angeles, CA, USA}
\email{samvizvary@gmail.com}

\author{Michael Bareian}
\affiliation{Department of Physics and Astronomy, University of California Los Angeles, Los Angeles, CA, USA}

\author{Andrei Derevianko}
\affiliation{Department of Physics, University of Nevada, Reno, NV, 89557, USA}

\author{Wesley C. Campbell}
\affiliation{Department of Physics and Astronomy, University of California Los Angeles, Los Angeles, CA, USA}
\affiliation{Challenge Institute for Quantum Computation, University of California Los Angeles, Los Angeles, CA, USA}
\affiliation{Center for Quantum Science and Engineering, University of California Los Angeles, Los Angeles, CA, USA}
\author{Eric R. Hudson}
\affiliation{Department of Physics and Astronomy, University of California Los Angeles, Los Angeles, CA, USA}
\affiliation{Challenge Institute for Quantum Computation, University of California Los Angeles, Los Angeles, CA, USA}
\affiliation{Center for Quantum Science and Engineering, University of California Los Angeles, Los Angeles, CA, USA}

\date{\today} 
\begin{abstract}
The $omg$ protocol is a promising paradigm
that uses multiple, application-specific qubit subspaces within the Hilbert space of each single atom during quantum information processing. 
A key assumption for $omg$ operation is that a subspace can be accessed independently without deleterious effects on information stored in other subspaces.
We find that intensity noise during laser-based quantum gates in one subspace can cause decoherence in other subspaces, potentially complicating $omg$ operation. 
We show, however, that a magnetic-field-induced vector light shift can be used to eliminate this source of decoherence.
As this technique requires simply choosing a certain, magnetic field dependent, polarization for the gate lasers it is straightforward to implement and potentially helpful for $omg$ based quantum technology.

\end{abstract}

\maketitle

The highest fidelity state-preparation and measurement (SPAM) operations~\cite{An2022} as well as single- and two-qubit gates~\cite{Ballance2016,Gaebler2016} have all been achieved using trapped atomic ions. 
This performance is realized in part thanks to the high degree of isolation of trapped ion qubits from their environment as compared to other technologies.
This isolation necessitates that the \emph{entropic} (or open-channel) operations used to prepare a qubit, \textit{i.e.}\ motional cooling and state preparation, are introduced deliberately, typically via laser light that is resonant with atomic electronic transitions. As light from these lasers can be extinguished as needed, they provide a strong but severable link to the environmental bath. 

However, if cooling is needed as part of a given quantum operation, due to, \textit{e.g.}, heating during a long algorithm~\cite{sun2023quantum} or due to shuttling in the QCCD architecture~\cite{Moses2023}, laser cooling cannot be performed directly since the strong coupling of the process to the atomic internal states scrambles any quantum information hosted in the atom. 
To overcome this limitation, some ion-based quantum processors simultaneously use two species of atoms~\cite{Moses2023}. 
One species, the logic ion, is used to host and process the quantum information, while the other ion, the coolant ion, is used only for cooling.

This `dual species' approach is  powerful, providing capabilities like mitigation of heating during transport and long algorithms, as well as facilitating mid-circuit measurement~\cite{Pino_2021}. 
It is not, however, without complication. 
Beyond requiring significantly more complicated laser and optical systems
, the mass difference of the two species, coupled with the pondermotive nature of an ion trap, leads to deleterious effects such as mode-decoupling \cite{Sosnova2021} and increased heating during transport~\cite{Shu2014,Blakestad2009}.
These necessitate complicated shuttling protocols, specific ion chain arrangements, and significant recooling time after transport \cite{Moses2023}. 

A recent proposal, dubbed the \emph{omg} protocol, has described how to achieve dual-species functionality using a single atomic species~\cite{Allcock2021}. 
The \emph{omg} protocol leverages the three types of qubits; $o$ptical, $m$etastable, and $g$round, available in certain atomic ions species to host quantum information in different parts of the atomic Hilbert space.
Allowing operations to be performed on some ions, while other ions, occupying different part of the Hilbert space, are protected from the operation. 
As such, \emph{omg} appears to offer many advantages over the dual-species approach including reducing the number of trapped ions needed for a given algorithm, 
a reduction in laser and optical complexity, 
the elimination of the effects associated with mass mismatch, and the ability to flexibly define an ion's role. 
\begin{figure}[t]
\includegraphics[width=.45\textwidth]{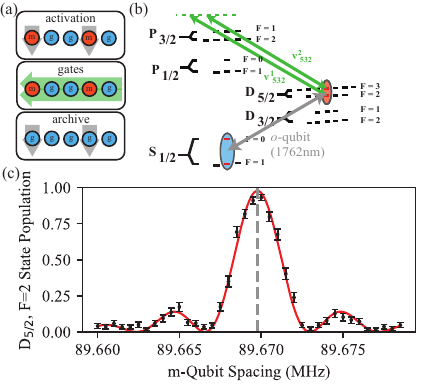}%
\caption{\label{fig:FIG_1_abc} (a) \{$g, m,g$\} $omg$ protocol for \barium. Cooling and readout are performed via $g$ qubits in the $^2$S$_{1/2}$ manifold. Activation of the $m$ qubit occurs via the $o$-qubit transition. Gates are done via global $m$-qubit beams. Archiving occurs with a transfer back to the $g$-qubit manifold. (b) Energy level diagram showing both the $g$ (blue) and $m$ (orange) manifolds in \barium. Both Raman gate beams ($\nu^1_{532}, \nu^2_{532}$) are split in frequency by the m-qubit spacing and blue detuned from the P$_{3/2}$ resonant transition. (c) Raman spectroscopy of $m$ qubit hyperfine splitting found to be 89.6697(4) MHz at 5.036(1)~G. }
\vspace{-10pt}
\end{figure}

However, while some of the basic components of the \emph{omg} protocol have been demonstrated, such as qubit initialization and conversion \cite{Yang2022}, many of its promised benefits remain unproven. 
One open question is the degree to which operations can be performed on one part of the Hilbert space without affecting others. 
Recent work has shown that laser cooling on the $g$-space does not noticeably affect coherent operations on the $m$-space \cite{Yang2022}.
However, laser cooling requires relatively low laser intensity as compared to the lasers used for single- and two- qubit gates.
Here, we show that the high-intensity operations necessary for laser-based gates in the $m$-space of \barium\ do indeed lead to detrimental effects, as the intensity instability of the gate lasers causes uncontrolled differential light shifts for qubits stored in the $g$-manifold. 
However, we find this effect can be completely mitigated by exploiting magnetic-field-induced hyperfine mixing to realize a non-zero, vector light shift between the clock state qubits that cancels the scalar and tensor shifts \cite{Porto2011}. 
Therefore, with the correct laser polarization, which we dub the `magic polarization',  the differential light shift of $g$-type qubits by the gate lasers is nulled, protecting information stored in $g$-type qubits from $m$-type qubit laser-based gate operations. 

Implementation of the $omg$ protocol in \barium\ utilizes the $\{g,m,g\}$ architecture -- here, the ordered triplet denotes the qubit space used for $\{\text{cooling, gates, storage}\}$~\cite{Allcock2021}. 
In this architecture, information is stored in the $g$ subspace and qubits are `activated' to the $m$ subspace for gates, as shown in Fig.~\ref{fig:FIG_1_abc}(a).
If the activation resolves single ions, the laser light used for gates can be applied globally, providing a dramatic simplification in systematic complexity. 
In this work, we define the $g$-type qubit as the $F=0$ and $F=1$ hyperfine zero-field clock (\textit{i.e.}\ $m_F = 0$) states of the \barium\ $^2$S$_{1/2}$ manifold, while the $m$-type qubit is defined on the the $F=2$ and $F=3$ hyperfine zero-field clock states of the $^2$D$_{5/2}$ manifold.  
Laser cooling, state preparation, and state readout are performed via the $g$ subspace as detailed in Ref.~\cite{Christensen20}. 
After high fidelity state preparation of the $F = 1, m_F = 0$ $g$-qubit state, transfer to the $m$-type qubit 
is accomplished via a heralded coherent operation on the $o$-qubit transition at 1762~nm to the $F = 3, m_F = 0$ state of the $^2$D$_{5/2}$ manifold (see SI).
Readout of the $m$ qubit is performed by a shelving operation 
that involves the simultaneous application of lasers at 1762~nm and 493~nm to transfer the target $m$-state population to the $^2$D$_{3/2}$ manifold, where it can be detected via resonant fluorescence (see SI).
A continuous wave laser at 532~nm is directed through acousto-optical modulators to generate the beams for stimulated Raman $m$-type qubit gates. An electro-optical modulator in the beam path, allows the same laser to also perform operations on the $g$-type qubit; while not required by the $\{g,m,g\}$ architecture the ability to use the same laser path to perform $g$ and $m$ qubit gates adds significant flexibility to the system. 

Because high-resolution spectroscopy of \barium\ is not yet complete, before performing $m$ qubit gates it is necessary to first measure the hyperfine splitting of the $5d6s$~$^2$D$_{5/2}$ state that hosts the $m$ qubit. We perform this measurement via a stimulated Raman spectroscopy on the $F = 3, m_F = 0 \leftrightarrow F = 2, m_F = 0$ transition at $B = 5.036(1)$~G, as shown in Fig. \ref{fig:FIG_1_abc}(b,c).
From this measurement, we report the first high-resolution measurement of the \barium\ $5d6s$~$^2$D$_{5/2}$  hyperfine constant as $A_{5/2} = 29.7565(1)_{sys}$ MHz, with a statistical uncertainty of 40 Hz. 
Using this measurement, we use a single $m$-qubit rotation to measure the intensity stability limited, $m$-qubit coherence time to be $2.6(9)$~ms (see SI).

\begin{figure}[t]
\includegraphics[width=.47\textwidth]{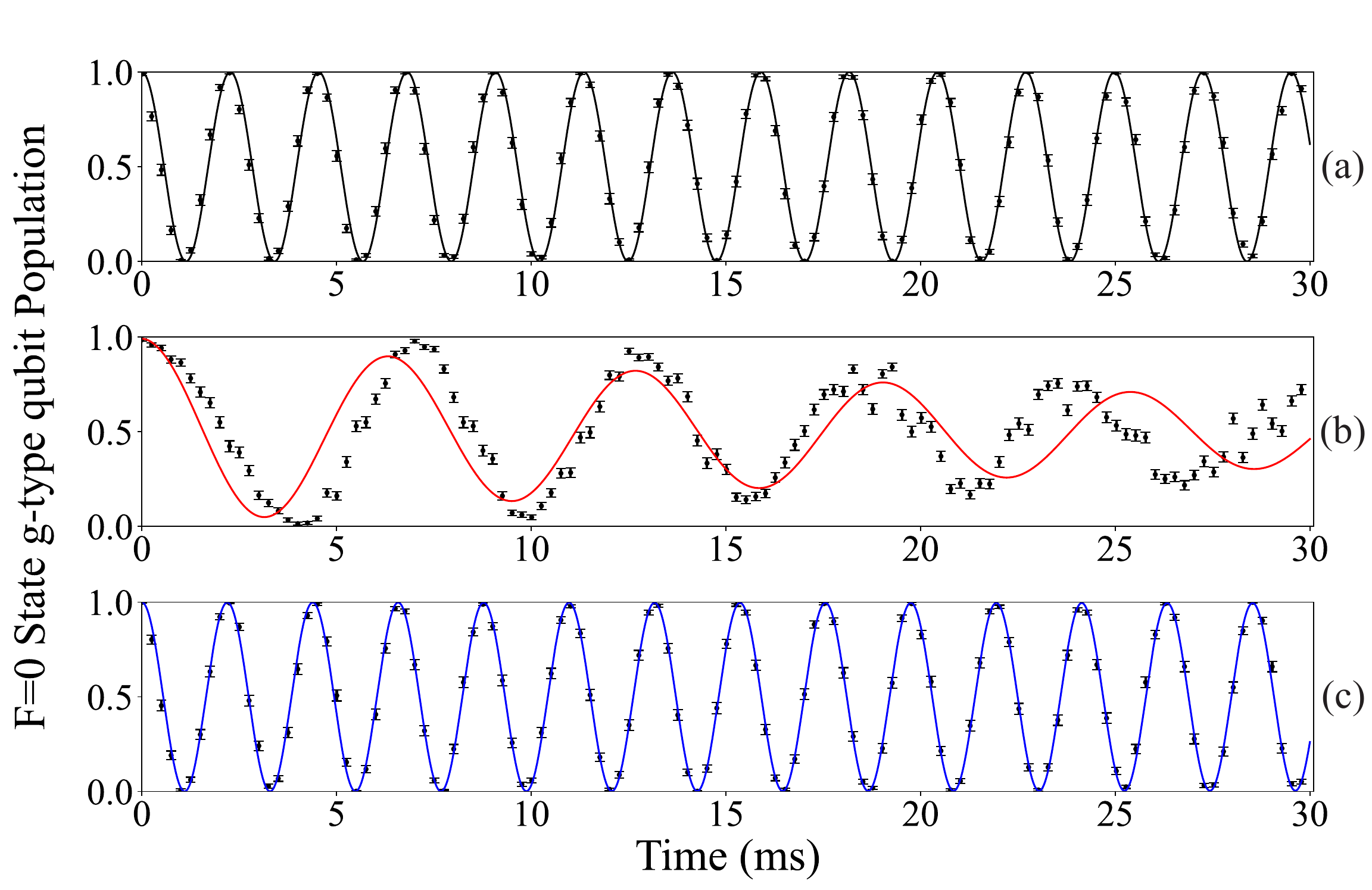}
\caption{\label{fig:Ramsey Curves}Population of the $g$-qubit $F = 0$ state under a detuned Ramsey sequences at 5.00(1) G with (a) no 532~nm gate laser illumination, (b) 532~nm illumination at a non-magic polarization and (c) 532~nm illumination at the magic polarization. }
\vspace{-5pt}
\end{figure}

\begin{figure*}[t]
\includegraphics[width=\textwidth]{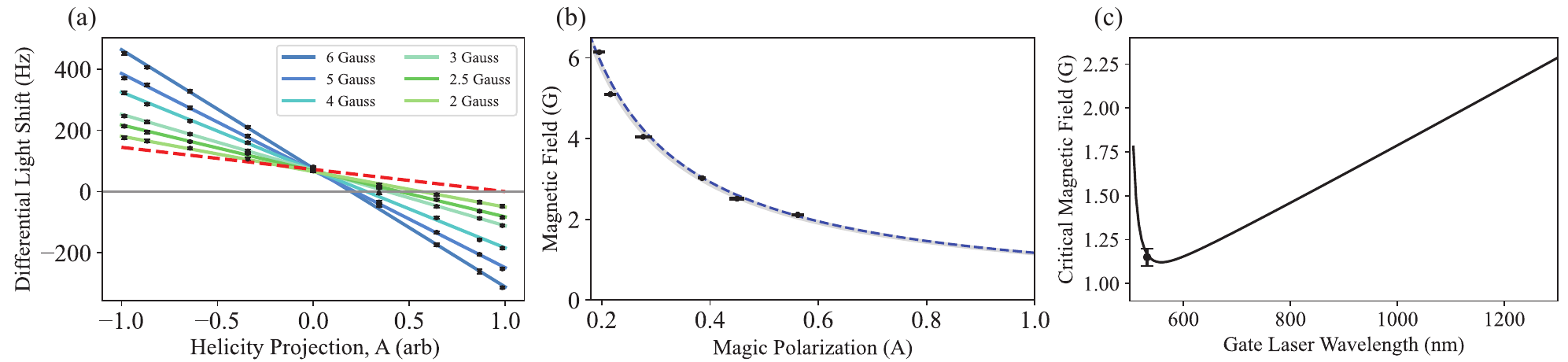}
    \caption{(a) Differential light shift of $g$ qubit spacing as a function of the helicity projection, $A$. 
    The $A$ value corresponding to the zero crossing for each magnetic field is the `magic conditions' where the differential shift is zero. 
    The red dashed line represents the theoretical curve for the critical field, B$_c$, found here. 
    (b) The applied magnetic field versus magic polarization determined from fits of the `magic conditions' in Fig. \ref{fig:b_field_measurments}(a). 
    The data is plotted as black dots with error bars and the grey shaded area is a 68\% confidence interval of the ODR fit.
    The critical field, B$_{c} = 1.15(3)$~G, is found at A=1 from the fit. 
    The dotted blue line represents an \emph{ab initio} theoretical calculation giving a value of B$_c$ in \barium of 1.17 G. 
    (c) Calculated critical magnetic field as a function of gate laser wavelenghth. The data point at 532 nm represents the value measured here.}
        \label{fig:b_field_measurments}
    \vspace{-10pt}
\end{figure*}

To ascertain the effect of the light used for these $m$-type gates on a qubit stored in the $g$ subspace, we perform a detuned Ramsey sequence on a $g$ qubit, using microwaves, with and without illumination by the laser beams used in the $m$-qubit gate. 
As seen in Fig.~\ref{fig:Ramsey Curves} for $\sigma^-$ polarization and a magnetic field of $B = 5.00(1)$~G, a fit of a simple decaying oscillation is shown as a guide to the eye and suggests a coherence time of approximately 30~ms. 

While this effect could be somewhat mitigated with technical improvements and dynamical decoupling, it would be beneficial if the differential light shift of the $g$-qubit states could simply be eliminated. 
To understand how this is possible, it is useful to write the light shift of an atomic state due to a laser at frequency $\omega_L$ and polarization $\hat{\epsilon}_L$ as $\Delta E_{F,m_F} \propto 
\alpha_F(m_F,\omega_L,\hat{\epsilon}_L) I_L$, where $\alpha_F$ is the dynamic polarizability and $I_L$ is the laser intensity.

The dynamic polarizability can be decomposed as~\cite{Manakov1986}:
\begin{align}
    \alpha_F(m_F,&\,\omega_L,\hat{\epsilon}_L) = \alpha_F^{s}(\omega_L) + A\,\frac{m_F}{2F}\,\alpha_F^{a}(\omega_L) \nonumber\\
    &+ \frac{1}{2}(3|\hat{\epsilon}_L\cdot\hat{z}|^2 - 1)\frac{3m_F^2 - F(F+1)}{F(2F-1)}\alpha_F^{t}(\omega_L),
    \label{eqn:alpha}
\end{align}
where the $m_F$-independent $\alpha^{s}_F$, $\alpha^{a}_F$, and $\alpha^{t}_F$ are the scalar, vector (axial), and tensor dynamic polarizabilities, respectively, $A \equiv -i(\hat{\epsilon}_L^\ast \times\hat{\epsilon}_L)\cdot\hat{z} $ is the helicity projection of circular polarization and $\hat{z}$ is the quantization axis. 
At first glance, since the vector and tensor polarizabilities are non-zero only for $F\geq1/2$ and $F\geq1$, respectively, and the $g$-qubit states have $m_F = 0$ it would appear the only opportunity for nulling the differential light shift between the qubit states is via controlling the direction of linear polarization relative to the magnetic field to tune the tensor shift contribution.
However, the magnitude of the tensor contribution is not large enough to allow matching the light shift of the two qubit states.
Fortunately, in a non-zero magnetic field Eq.~\eqref{eqn:alpha} is only approximate as the magnetic field breaks rotational invariance and $F$ is no longer a good quantum number. 
The result is a non-zero vector light shift contribution even for $m_F = 0$~\cite{Porto2011}, which can be used to match the dynamic polarizabilities of the two clock states and null the differential light shift. 

This can be intuitively understood by considering the light shift of the $g$-qubit states under illumination by a laser with $\sigma_+$ polarization. 
The two states couple primarily to the $F = m_F = 1$ state of the $^2\mathrm{P}_{1/2}$ manifold (see e.g. Fig. \ref{fig:FIG_1_abc}(b)).
At zero magnetic field, the $F = 0$ $g$ qubit state experiences a larger light shift than the $F = 1$ $g$ qubit state because it is closer to the $^2\mathrm{P}_{1/2}$ manifold. 
Therefore, the $g$ qubit has a negative differential light shift. 
In a non-zero magnetic field, the two $g$-qubit states are mixed and the coupling is modified such that the light shift now depends on the degree of helicity projection. 

To quantify this effect, we measure the differential light shift of the $g$ qubit using a Ramsey sequence (see SI) as a function of $A$ and $B\hat{z}$, under the condition that the angle between $\hat{k}$ and $\hat{z}$ is $\theta_{kz} = 180.0(8)^\circ$. 
As seen in Fig.~\ref{fig:b_field_measurments}, the $g$ qubit exhibits a clear, magnetic-field dependent vector light shift.
Interestingly, above a certain magnetic field this vector light shift is large enough to completely counteract the difference in the scalar light shifts.
By using gate lasers operating at a polarization where the differential light shift is zero, which we dub a magic polarization, the $g$-qubit splitting becomes completely insensitive to the $m$-qubit gate lasers. 
The magnetic field required for a given magic polarization is found by finding the zero crossing of the fitted line in Fig.~\ref{fig:b_field_measurments}(a) and plotted in Fig.~\ref{fig:b_field_measurments}(b) as black points, where the error bars are 68.5\% confidence intervals from the fit. 
The minimum magnetic field necessary for cancellation of the $g$-qubit light shift, which we dub the critical magnetic field $B_c$, is achieved for $A = 1$ ($\sigma_+$ polarization). 
However, due to coherent population trapping effects \cite{Gray1978}, our Doppler cooling efficiency suffers below a magnetic field of $\sim 2$~G, and therefore we cannot probe the critical magnetic field directly. 
Instead, we fit the expected theoretical behavior to the data to extract $B_c$.

The dependence of the magic polarization on magnetic field is found by diagonalizing the optical Hamiltonian leading to Eq.~\eqref{eqn:alpha} in the presence of a magnetic field $B$~\cite{Der10Bmagic}.
The result is a vector light shift term that is proportional to $B$ leading to the relation:
\begin{equation}
A \approx-\frac{\hbar \omega_q}{\mu_B B} \sqrt{\frac{2I+1}{2I(2I+2)}} \frac{\alpha_{F'}^{s}\left(\omega_L\right) - \alpha_{F}^{s}\left(\omega_L\right)} {\alpha_{F}^{a}\left(\omega_L\right)} \approx \frac{B_c}{B}\,, 
\label{A_crit}
\end{equation}
where we distinguish between the polarizabilities of the two ($F'=0$ and $F=1$) qubit states separated by energy $\hbar \omega_q$.
Fitting this expression to the data in Fig.~\ref{fig:b_field_measurments}(b)(the grey shaded region), yields a critical magnetic field for \barium\ $g$ qubits of $B_c = 1.15(3)$~G at 532~nm. 

Calculation of the expected behavior requires a third-order hyperfine-interaction (HFI) mediated polarizabilities treatment~\cite{Der10Bmagic} be used in Eq.~\eqref{A_crit} as the two scalar polarizabilities cancel out otherwise.
This partially stems from the fact that the total electron angular momentum $J$ is mixed by the atomic hyperfine interaction, leading to differing transition dipole moments for the hyperfine qubit states to the same fine-structure state. 
Using a combination of empirical data for low-lying electronic states and {\em ab initio} relativistic RPA+BO many-body method~\cite{tan2023precision,Der10Bmagic,XiaTanDer2023-CsNucSpinTranPol}, we numerical evaluate the polarizabilities in Eq.~\eqref{A_crit}. 
The results of the numerical calculation are plotted as the dashed blue line in Fig.~\ref{fig:b_field_measurments}(b), and correspond to a  predicted critical magnetic field of at 532~nm of $B_c =  1.17$~G in agreement with the measurement.

The HFI mediated polarizability calculation has also been performed as a function of gate laser wavelength and the predicted $B_c$ is plotted in Fig.~\ref{fig:b_field_measurments}(c) alongside the data.
The rise in the $B_c$ with large detunings is due to interference from the counter-rotating terms, which work to lessen the induced vector light shift.
This difference can be substantial and clearly shows the need for high-level calculations. 

To demonstrate the utility of magic polarization for $omg$ operation, we repeat the measurements of Fig.~\ref{fig:Ramsey Curves} at $B = 5.00(1)$~G, but with the polarization chosen to be magic ($A = 0.212$). 
As can be seen in Fig.~\ref{fig:Ramsey Curves}, the decoherence due to the light shift of the $g$ qubit by the $m$-qubit gate laser is mitigated, leading to a $\sim$100-fold increase in coherence time.
In fact, the $g$-qubit coherence time of ($\tau = 3.6(2.5)$~s) is consistent with the performance observed when the ion is not illuminated by the gate laser (Fig.~\ref{fig:Ramsey Curves}) despite being exposed to a laser field with an intensity of $\sim 100$~MW/m$^2$. 

For single and two-qubit gates, $A = 1$ polarization is preferred as it yields the largest Rabi rate and therefore minimizes spontaneous Raman scattering error~\cite{Moore2023, Boguslawski2023}. 
Thus, to harness the benefits of this magic condition, the applied magnetic field should be at the critical point. 
However, at 532~nm the critical magnetic field for \barium\ is well below 2~G, which is the minimum magnetic field we require for efficient Doppler cooling. 
Luckily, as the gate laser is further detuned to longer wavelengths the critical field grows, due to interference from the emission first term in the light shift, as shown in Fig. \ref{fig:b_field_measurments}(c).
 
The use of a longer wavelength gate laser, while requiring moderately more power, has the added benefit of a lower Raman scattering rate. 
For example, at 1130~nm $B_c \approx 2$~G and the two-qubit gate infidelity due to Raman scattering is expected to be $< 10^{-5}$ ~\cite{Boguslawski2023}.

In summary, we demonstrate the first preparation, operation, and readout of the $m$ qubit in \barium\ and report the most accurate measurement to date of the $5d6s$ $^2$D$_{5/2}$ hyperfine splitting of $A_{5/2} = 29.75650(10)_{sys}(4)_{stat}$ MHz.
Using these tools, we examine the cross-talk of $omg$ laser-based gates between qubit subspaces and find that gate operations on $m$ qubits lead to decoherence of $g$ qubits.
This decoherence results from the $g$ qubit differential light shift by the gate laser. 
However, utilizing a magnetic-field-induced vector light shift we show how the light shift between these zero-field clock state qubits can be matched using a magic polarization. 
We measure the critical magnetic field required to realize a magic polarization in \barium\ and compare it to high-level atomic structure calculations and find good agreement. 
Finally, using the magic polarization to null the differential light shift of the $g$-qubit states, we demonstrate protection of quantum information stored in the $g$ qubit from $m$-qubit laser-based operations.

\begin{acknowledgments}
This work was supported in part by the  the Army Research Office grant W911NF-20-1-0037 and National Science Foundation grants PHY-2207546 and OMA-2016245.
\end{acknowledgments}

\bibliographystyle{apsrev4-1}
\bibliography{ref,library-apd}

\begin{thebibliography}{20}%
\makeatletter
\providecommand \@ifxundefined [1]{%
 \@ifx{#1\undefined}
}%
\providecommand \@ifnum [1]{%
 \ifnum #1\expandafter \@firstoftwo
 \else \expandafter \@secondoftwo
 \fi
}%
\providecommand \@ifx [1]{%
 \ifx #1\expandafter \@firstoftwo
 \else \expandafter \@secondoftwo
 \fi
}%
\providecommand \natexlab [1]{#1}%
\providecommand \enquote  [1]{``#1''}%
\providecommand \bibnamefont  [1]{#1}%
\providecommand \bibfnamefont [1]{#1}%
\providecommand \citenamefont [1]{#1}%
\providecommand \href@noop [0]{\@secondoftwo}%
\providecommand \href [0]{\begingroup \@sanitize@url \@href}%
\providecommand \@href[1]{\@@startlink{#1}\@@href}%
\providecommand \@@href[1]{\endgroup#1\@@endlink}%
\providecommand \@sanitize@url [0]{\catcode `\\12\catcode `\$12\catcode
  `\&12\catcode `\#12\catcode `\^12\catcode `\_12\catcode `\%12\relax}%
\providecommand \@@startlink[1]{}%
\providecommand \@@endlink[0]{}%
\providecommand \url  [0]{\begingroup\@sanitize@url \@url }%
\providecommand \@url [1]{\endgroup\@href {#1}{\urlprefix }}%
\providecommand \urlprefix  [0]{URL }%
\providecommand \Eprint [0]{\href }%
\providecommand \doibase [0]{http://dx.doi.org/}%
\providecommand \selectlanguage [0]{\@gobble}%
\providecommand \bibinfo  [0]{\@secondoftwo}%
\providecommand \bibfield  [0]{\@secondoftwo}%
\providecommand \translation [1]{[#1]}%
\providecommand \BibitemOpen [0]{}%
\providecommand \bibitemStop [0]{}%
\providecommand \bibitemNoStop [0]{.\EOS\space}%
\providecommand \EOS [0]{\spacefactor3000\relax}%
\providecommand \BibitemShut  [1]{\csname bibitem#1\endcsname}%
\let\auto@bib@innerbib\@empty
\bibitem [{\citenamefont {An}\ \emph {et~al.}(2022)\citenamefont {An},
  \citenamefont {Ransford}, \citenamefont {Schaffer}, \citenamefont {Sletten},
  \citenamefont {Gaebler}, \citenamefont {Hostetter},\ and\ \citenamefont
  {Vittorini}}]{An2022}%
  \BibitemOpen
  \bibfield  {author} {\bibinfo {author} {\bibfnamefont {F.~A.}\ \bibnamefont
  {An}}, \bibinfo {author} {\bibfnamefont {A.}~\bibnamefont {Ransford}},
  \bibinfo {author} {\bibfnamefont {A.}~\bibnamefont {Schaffer}}, \bibinfo
  {author} {\bibfnamefont {L.~R.}\ \bibnamefont {Sletten}}, \bibinfo {author}
  {\bibfnamefont {J.}~\bibnamefont {Gaebler}}, \bibinfo {author} {\bibfnamefont
  {J.}~\bibnamefont {Hostetter}}, \ and\ \bibinfo {author} {\bibfnamefont
  {G.}~\bibnamefont {Vittorini}},\ }\href {\doibase
  10.1103/PhysRevLett.129.130501} {\bibfield  {journal} {\bibinfo  {journal}
  {Phys. Rev. Lett.}\ }\textbf {\bibinfo {volume} {129}},\ \bibinfo {pages}
  {130501} (\bibinfo {year} {2022})}\BibitemShut {NoStop}%
\bibitem [{\citenamefont {Ballance}\ \emph {et~al.}(2016)\citenamefont
  {Ballance}, \citenamefont {Harty}, \citenamefont {Linke}, \citenamefont
  {Sepiol},\ and\ \citenamefont {Lucas}}]{Ballance2016}%
  \BibitemOpen
  \bibfield  {author} {\bibinfo {author} {\bibfnamefont {C.~J.}\ \bibnamefont
  {Ballance}}, \bibinfo {author} {\bibfnamefont {T.~P.}\ \bibnamefont {Harty}},
  \bibinfo {author} {\bibfnamefont {N.~M.}\ \bibnamefont {Linke}}, \bibinfo
  {author} {\bibfnamefont {M.~A.}\ \bibnamefont {Sepiol}}, \ and\ \bibinfo
  {author} {\bibfnamefont {D.~M.}\ \bibnamefont {Lucas}},\ }\href {\doibase
  10.1103/PhysRevLett.117.060504} {\bibfield  {journal} {\bibinfo  {journal}
  {Phys. Rev. Lett.}\ }\textbf {\bibinfo {volume} {117}},\ \bibinfo {pages}
  {060504} (\bibinfo {year} {2016})}\BibitemShut {NoStop}%
\bibitem [{\citenamefont {Gaebler}\ \emph {et~al.}(2016)\citenamefont
  {Gaebler}, \citenamefont {Tan}, \citenamefont {Lin}, \citenamefont {Wan},
  \citenamefont {Bowler}, \citenamefont {Keith}, \citenamefont {Glancy},
  \citenamefont {Coakley}, \citenamefont {Knill}, \citenamefont {Leibfried},\
  and\ \citenamefont {Wineland}}]{Gaebler2016}%
  \BibitemOpen
  \bibfield  {author} {\bibinfo {author} {\bibfnamefont {J.~P.}\ \bibnamefont
  {Gaebler}}, \bibinfo {author} {\bibfnamefont {T.~R.}\ \bibnamefont {Tan}},
  \bibinfo {author} {\bibfnamefont {Y.}~\bibnamefont {Lin}}, \bibinfo {author}
  {\bibfnamefont {Y.}~\bibnamefont {Wan}}, \bibinfo {author} {\bibfnamefont
  {R.}~\bibnamefont {Bowler}}, \bibinfo {author} {\bibfnamefont {A.~C.}\
  \bibnamefont {Keith}}, \bibinfo {author} {\bibfnamefont {S.}~\bibnamefont
  {Glancy}}, \bibinfo {author} {\bibfnamefont {K.}~\bibnamefont {Coakley}},
  \bibinfo {author} {\bibfnamefont {E.}~\bibnamefont {Knill}}, \bibinfo
  {author} {\bibfnamefont {D.}~\bibnamefont {Leibfried}}, \ and\ \bibinfo
  {author} {\bibfnamefont {D.~J.}\ \bibnamefont {Wineland}},\ }\href {\doibase
  10.1103/PhysRevLett.117.060505} {\bibfield  {journal} {\bibinfo  {journal}
  {Phys. Rev. Lett.}\ }\textbf {\bibinfo {volume} {117}},\ \bibinfo {pages}
  {060505} (\bibinfo {year} {2016})}\BibitemShut {NoStop}%
\bibitem [{\citenamefont {Sun}\ \emph {et~al.}(2023)\citenamefont {Sun},
  \citenamefont {Fang}, \citenamefont {Kang}, \citenamefont {Zhang},
  \citenamefont {Zhang}, \citenamefont {Beratan}, \citenamefont {Brown},\ and\
  \citenamefont {Kim}}]{sun2023quantum}%
  \BibitemOpen
  \bibfield  {author} {\bibinfo {author} {\bibfnamefont {K.}~\bibnamefont
  {Sun}}, \bibinfo {author} {\bibfnamefont {C.}~\bibnamefont {Fang}}, \bibinfo
  {author} {\bibfnamefont {M.}~\bibnamefont {Kang}}, \bibinfo {author}
  {\bibfnamefont {Z.}~\bibnamefont {Zhang}}, \bibinfo {author} {\bibfnamefont
  {P.}~\bibnamefont {Zhang}}, \bibinfo {author} {\bibfnamefont {D.~N.}\
  \bibnamefont {Beratan}}, \bibinfo {author} {\bibfnamefont {K.~R.}\
  \bibnamefont {Brown}}, \ and\ \bibinfo {author} {\bibfnamefont
  {J.}~\bibnamefont {Kim}},\ }\href@noop {} {\enquote {\bibinfo {title}
  {Quantum simulation of polarized light-induced electron transfer with a
  trapped-ion qutrit system},}\ } (\bibinfo {year} {2023}),\ \Eprint
  {http://arxiv.org/abs/2304.12247} {arXiv:2304.12247 [quant-ph]} \BibitemShut
  {NoStop}%
\bibitem [{\citenamefont {Moses}\ \emph {et~al.}(2023)\citenamefont {Moses}
  \emph {et~al.}}]{Moses2023}%
  \BibitemOpen
  \bibfield  {author} {\bibinfo {author} {\bibfnamefont {S.~A.}\ \bibnamefont
  {Moses}} \emph {et~al.},\ }\href@noop {} {\enquote {\bibinfo {title} {A race
  track trapped-ion quantum processor},}\ } (\bibinfo {year} {2023}),\ \Eprint
  {http://arxiv.org/abs/2305.03828} {arXiv:2305.03828 [quant-ph]} \BibitemShut
  {NoStop}%
\bibitem [{\citenamefont {Pino}\ \emph {et~al.}(2021)\citenamefont {Pino},
  \citenamefont {Dreiling}, \citenamefont {Figgatt}, \citenamefont {Gaebler},
  \citenamefont {Moses}, \citenamefont {Allman}, \citenamefont {Baldwin},
  \citenamefont {Foss-Feig}, \citenamefont {Hayes}, \citenamefont {Mayer},
  \citenamefont {Ryan-Anderson},\ and\ \citenamefont {Neyenhuis}}]{Pino_2021}%
  \BibitemOpen
  \bibfield  {author} {\bibinfo {author} {\bibfnamefont {J.~M.}\ \bibnamefont
  {Pino}}, \bibinfo {author} {\bibfnamefont {J.~M.}\ \bibnamefont {Dreiling}},
  \bibinfo {author} {\bibfnamefont {C.}~\bibnamefont {Figgatt}}, \bibinfo
  {author} {\bibfnamefont {J.~P.}\ \bibnamefont {Gaebler}}, \bibinfo {author}
  {\bibfnamefont {S.~A.}\ \bibnamefont {Moses}}, \bibinfo {author}
  {\bibfnamefont {M.~S.}\ \bibnamefont {Allman}}, \bibinfo {author}
  {\bibfnamefont {C.~H.}\ \bibnamefont {Baldwin}}, \bibinfo {author}
  {\bibfnamefont {M.}~\bibnamefont {Foss-Feig}}, \bibinfo {author}
  {\bibfnamefont {D.}~\bibnamefont {Hayes}}, \bibinfo {author} {\bibfnamefont
  {K.}~\bibnamefont {Mayer}}, \bibinfo {author} {\bibfnamefont
  {C.}~\bibnamefont {Ryan-Anderson}}, \ and\ \bibinfo {author} {\bibfnamefont
  {B.}~\bibnamefont {Neyenhuis}},\ }\href {\doibase 10.1038/s41586-021-03318-4}
  {\bibfield  {journal} {\bibinfo  {journal} {Nature}\ }\textbf {\bibinfo
  {volume} {592}},\ \bibinfo {pages} {209} (\bibinfo {year}
  {2021})}\BibitemShut {NoStop}%
\bibitem [{\citenamefont {Sosnova}\ \emph {et~al.}(2021)\citenamefont
  {Sosnova}, \citenamefont {Carter},\ and\ \citenamefont
  {Monroe}}]{Sosnova2021}%
  \BibitemOpen
  \bibfield  {author} {\bibinfo {author} {\bibfnamefont {K.}~\bibnamefont
  {Sosnova}}, \bibinfo {author} {\bibfnamefont {A.}~\bibnamefont {Carter}}, \
  and\ \bibinfo {author} {\bibfnamefont {C.}~\bibnamefont {Monroe}},\ }\href
  {\doibase 10.1103/PhysRevA.103.012610} {\bibfield  {journal} {\bibinfo
  {journal} {Phys. Rev. A}\ }\textbf {\bibinfo {volume} {103}},\ \bibinfo
  {pages} {012610} (\bibinfo {year} {2021})}\BibitemShut {NoStop}%
\bibitem [{\citenamefont {Shu}\ \emph {et~al.}(2014)\citenamefont {Shu},
  \citenamefont {Vittorini}, \citenamefont {Buikema}, \citenamefont {Nichols},
  \citenamefont {Volin}, \citenamefont {Stick},\ and\ \citenamefont
  {Brown}}]{Shu2014}%
  \BibitemOpen
  \bibfield  {author} {\bibinfo {author} {\bibfnamefont {G.}~\bibnamefont
  {Shu}}, \bibinfo {author} {\bibfnamefont {G.}~\bibnamefont {Vittorini}},
  \bibinfo {author} {\bibfnamefont {A.}~\bibnamefont {Buikema}}, \bibinfo
  {author} {\bibfnamefont {C.~S.}\ \bibnamefont {Nichols}}, \bibinfo {author}
  {\bibfnamefont {C.}~\bibnamefont {Volin}}, \bibinfo {author} {\bibfnamefont
  {D.}~\bibnamefont {Stick}}, \ and\ \bibinfo {author} {\bibfnamefont {K.~R.}\
  \bibnamefont {Brown}},\ }\href {\doibase 10.1103/PhysRevA.89.062308}
  {\bibfield  {journal} {\bibinfo  {journal} {Phys. Rev. A}\ }\textbf {\bibinfo
  {volume} {89}},\ \bibinfo {pages} {062308} (\bibinfo {year}
  {2014})}\BibitemShut {NoStop}%
\bibitem [{\citenamefont {Blakestad}\ \emph {et~al.}(2009)\citenamefont
  {Blakestad}, \citenamefont {Ospelkaus}, \citenamefont {VanDevender},
  \citenamefont {Amini}, \citenamefont {Britton}, \citenamefont {Leibfried},\
  and\ \citenamefont {Wineland}}]{Blakestad2009}%
  \BibitemOpen
  \bibfield  {author} {\bibinfo {author} {\bibfnamefont {R.~B.}\ \bibnamefont
  {Blakestad}}, \bibinfo {author} {\bibfnamefont {C.}~\bibnamefont
  {Ospelkaus}}, \bibinfo {author} {\bibfnamefont {A.~P.}\ \bibnamefont
  {VanDevender}}, \bibinfo {author} {\bibfnamefont {J.~M.}\ \bibnamefont
  {Amini}}, \bibinfo {author} {\bibfnamefont {J.}~\bibnamefont {Britton}},
  \bibinfo {author} {\bibfnamefont {D.}~\bibnamefont {Leibfried}}, \ and\
  \bibinfo {author} {\bibfnamefont {D.~J.}\ \bibnamefont {Wineland}},\ }\href
  {\doibase 10.1103/PhysRevLett.102.153002} {\bibfield  {journal} {\bibinfo
  {journal} {Phys. Rev. Lett.}\ }\textbf {\bibinfo {volume} {102}},\ \bibinfo
  {pages} {153002} (\bibinfo {year} {2009})}\BibitemShut {NoStop}%
\bibitem [{\citenamefont {Allcock}\ \emph {et~al.}(2021)\citenamefont
  {Allcock}, \citenamefont {Campbell}, \citenamefont {Chiaverini},
  \citenamefont {Chuang}, \citenamefont {Hudson}, \citenamefont {Moore},
  \citenamefont {Ransford}, \citenamefont {Roman}, \citenamefont {Sage},\ and\
  \citenamefont {Wineland}}]{Allcock2021}%
  \BibitemOpen
  \bibfield  {author} {\bibinfo {author} {\bibfnamefont {D.~T.~C.}\
  \bibnamefont {Allcock}}, \bibinfo {author} {\bibfnamefont {W.~C.}\
  \bibnamefont {Campbell}}, \bibinfo {author} {\bibfnamefont {J.}~\bibnamefont
  {Chiaverini}}, \bibinfo {author} {\bibfnamefont {I.~L.}\ \bibnamefont
  {Chuang}}, \bibinfo {author} {\bibfnamefont {E.~R.}\ \bibnamefont {Hudson}},
  \bibinfo {author} {\bibfnamefont {I.~D.}\ \bibnamefont {Moore}}, \bibinfo
  {author} {\bibfnamefont {A.}~\bibnamefont {Ransford}}, \bibinfo {author}
  {\bibfnamefont {C.}~\bibnamefont {Roman}}, \bibinfo {author} {\bibfnamefont
  {J.~M.}\ \bibnamefont {Sage}}, \ and\ \bibinfo {author} {\bibfnamefont
  {D.~J.}\ \bibnamefont {Wineland}},\ }\href
  {https://doi.org/10.1063/5.0069544} {\bibfield  {journal} {\bibinfo
  {journal} {Appl. Phys. Lett.}\ }\textbf {\bibinfo {volume} {119}},\ \bibinfo
  {pages} {214002} (\bibinfo {year} {2021})}\BibitemShut {NoStop}%
\bibitem [{\citenamefont {Yang}\ \emph {et~al.}(2022)\citenamefont {Yang},
  \citenamefont {Ma}, \citenamefont {Wu}, \citenamefont {Wang}, \citenamefont
  {Cao}, \citenamefont {Guo}, \citenamefont {Huang}, \citenamefont {Feng},
  \citenamefont {Zhou},\ and\ \citenamefont {Duan}}]{Yang2022}%
  \BibitemOpen
  \bibfield  {author} {\bibinfo {author} {\bibfnamefont {H.~X.}\ \bibnamefont
  {Yang}}, \bibinfo {author} {\bibfnamefont {J.~Y.}\ \bibnamefont {Ma}},
  \bibinfo {author} {\bibfnamefont {Y.~K.}\ \bibnamefont {Wu}}, \bibinfo
  {author} {\bibfnamefont {Y.}~\bibnamefont {Wang}}, \bibinfo {author}
  {\bibfnamefont {M.~M.}\ \bibnamefont {Cao}}, \bibinfo {author} {\bibfnamefont
  {W.-X.}\ \bibnamefont {Guo}}, \bibinfo {author} {\bibfnamefont {Y.~Y.}\
  \bibnamefont {Huang}}, \bibinfo {author} {\bibfnamefont {L.}~\bibnamefont
  {Feng}}, \bibinfo {author} {\bibfnamefont {Z.~C.}\ \bibnamefont {Zhou}}, \
  and\ \bibinfo {author} {\bibfnamefont {L.~M.}\ \bibnamefont {Duan}},\ }\href
  {\doibase 10.1038/s41567-022-01661-5} {\bibfield  {journal} {\bibinfo
  {journal} {Nat. Phys.}\ }\textbf {\bibinfo {volume} {18}},\ \bibinfo {pages}
  {1058–1061} (\bibinfo {year} {2022})}\BibitemShut {NoStop}%
\bibitem [{\citenamefont {Chicireanu}\ \emph {et~al.}(2011)\citenamefont
  {Chicireanu}, \citenamefont {Nelson}, \citenamefont {Olmschenk},
  \citenamefont {Lundblad}, \citenamefont {Derevianko},\ and\ \citenamefont
  {Porto}}]{Porto2011}%
  \BibitemOpen
  \bibfield  {author} {\bibinfo {author} {\bibfnamefont {R.}~\bibnamefont
  {Chicireanu}}, \bibinfo {author} {\bibfnamefont {K.~D.}\ \bibnamefont
  {Nelson}}, \bibinfo {author} {\bibfnamefont {S.}~\bibnamefont {Olmschenk}},
  \bibinfo {author} {\bibfnamefont {N.}~\bibnamefont {Lundblad}}, \bibinfo
  {author} {\bibfnamefont {A.}~\bibnamefont {Derevianko}}, \ and\ \bibinfo
  {author} {\bibfnamefont {J.~V.}\ \bibnamefont {Porto}},\ }\href {\doibase
  10.1103/PhysRevLett.106.063002} {\bibfield  {journal} {\bibinfo  {journal}
  {Phys. Rev. Lett.}\ }\textbf {\bibinfo {volume} {106}},\ \bibinfo {pages}
  {063002} (\bibinfo {year} {2011})}\BibitemShut {NoStop}%
\bibitem [{\citenamefont {Christensen}\ \emph {et~al.}(2020)\citenamefont
  {Christensen}, \citenamefont {Hucul}, \citenamefont {Campbell},\ and\
  \citenamefont {Hudson}}]{Christensen20}%
  \BibitemOpen
  \bibfield  {author} {\bibinfo {author} {\bibfnamefont {J.~E.}\ \bibnamefont
  {Christensen}}, \bibinfo {author} {\bibfnamefont {D.}~\bibnamefont {Hucul}},
  \bibinfo {author} {\bibfnamefont {W.~C.}\ \bibnamefont {Campbell}}, \ and\
  \bibinfo {author} {\bibfnamefont {E.~R.}\ \bibnamefont {Hudson}},\ }\href
  {\doibase 10.1038/s41534-020-0265-5} {\bibfield  {journal} {\bibinfo
  {journal} {npj Quantum Inf}\ }\textbf {\bibinfo {volume} {6}},\ \bibinfo
  {pages} {35} (\bibinfo {year} {2020})}\BibitemShut {NoStop}%
\bibitem [{\citenamefont {N.L.~Manakov}(1986)}]{Manakov1986}%
  \BibitemOpen
  \bibfield  {author} {\bibinfo {author} {\bibfnamefont {L.~R.}\ \bibnamefont
  {N.L.~Manakov}, \bibfnamefont {V.D.~Ovsiannikov}},\ }\href
  {https://doi.org/10.1016/S0370-1573(86)80001-1} {\bibfield  {journal}
  {\bibinfo  {journal} {Physics Reports}\ }\textbf {\bibinfo {volume} {141}},\
  \bibinfo {pages} {320} (\bibinfo {year} {1986})}\BibitemShut {NoStop}%
\bibitem [{\citenamefont {Gray}\ \emph {et~al.}(1978)\citenamefont {Gray},
  \citenamefont {Whitley},\ and\ \citenamefont {Stroud}}]{Gray1978}%
  \BibitemOpen
  \bibfield  {author} {\bibinfo {author} {\bibfnamefont {H.~R.}\ \bibnamefont
  {Gray}}, \bibinfo {author} {\bibfnamefont {R.~M.}\ \bibnamefont {Whitley}}, \
  and\ \bibinfo {author} {\bibfnamefont {C.~R.}\ \bibnamefont {Stroud}},\
  }\href {\doibase 10.1364/OL.3.000218} {\bibfield  {journal} {\bibinfo
  {journal} {Opt. Lett.}\ }\textbf {\bibinfo {volume} {3}},\ \bibinfo {pages}
  {218} (\bibinfo {year} {1978})}\BibitemShut {NoStop}%
\bibitem [{\citenamefont {Derevianko}(2010)}]{Der10Bmagic}%
  \BibitemOpen
  \bibfield  {author} {\bibinfo {author} {\bibfnamefont {A.}~\bibnamefont
  {Derevianko}},\ }\href {\doibase 10.1103/PhysRevA.81.051606} {\bibfield
  {journal} {\bibinfo  {journal} {Phys. Rev. A}\ }\textbf {\bibinfo {volume}
  {81}},\ \bibinfo {pages} {051606(R)} (\bibinfo {year} {2010})}\BibitemShut
  {NoStop}%
\bibitem [{\citenamefont {{Tran Tan}}\ and\ \citenamefont
  {Derevianko}(2023)}]{tan2023precision}%
  \BibitemOpen
  \bibfield  {author} {\bibinfo {author} {\bibfnamefont {H.~B.}\ \bibnamefont
  {{Tran Tan}}}\ and\ \bibinfo {author} {\bibfnamefont {A.}~\bibnamefont
  {Derevianko}},\ }\href {\doibase 10.1103/PhysRevA.107.042809} {\bibfield
  {journal} {\bibinfo  {journal} {Phys. Rev. A}\ }\textbf {\bibinfo {volume}
  {107}},\ \bibinfo {pages} {042809} (\bibinfo {year} {2023})}\BibitemShut
  {NoStop}%
\bibitem [{\citenamefont {Xiao}\ \emph {et~al.}(2023)\citenamefont {Xiao},
  \citenamefont {Tan},\ and\ \citenamefont
  {Derevianko}}]{XiaTanDer2023-CsNucSpinTranPol}%
  \BibitemOpen
  \bibfield  {author} {\bibinfo {author} {\bibfnamefont {D.}~\bibnamefont
  {Xiao}}, \bibinfo {author} {\bibfnamefont {H.~B.~T.}\ \bibnamefont {Tan}}, \
  and\ \bibinfo {author} {\bibfnamefont {A.}~\bibnamefont {Derevianko}},\
  }\href {\doibase 10.1103/PhysRevA.108.032805} {\bibfield  {journal} {\bibinfo
   {journal} {Phys. Rev. A}\ }\textbf {\bibinfo {volume} {108}},\ \bibinfo
  {pages} {032805} (\bibinfo {year} {2023})},\ \Eprint
  {http://arxiv.org/abs/2307.04272} {arXiv:2307.04272} \BibitemShut {NoStop}%
\bibitem [{\citenamefont {Moore}\ \emph {et~al.}(2023)\citenamefont {Moore},
  \citenamefont {Campbell}, \citenamefont {Hudson}, \citenamefont
  {Boguslawski}, \citenamefont {Wineland},\ and\ \citenamefont
  {Allcock}}]{Moore2023}%
  \BibitemOpen
  \bibfield  {author} {\bibinfo {author} {\bibfnamefont {I.~D.}\ \bibnamefont
  {Moore}}, \bibinfo {author} {\bibfnamefont {W.~C.}\ \bibnamefont {Campbell}},
  \bibinfo {author} {\bibfnamefont {E.~R.}\ \bibnamefont {Hudson}}, \bibinfo
  {author} {\bibfnamefont {M.~J.}\ \bibnamefont {Boguslawski}}, \bibinfo
  {author} {\bibfnamefont {D.~J.}\ \bibnamefont {Wineland}}, \ and\ \bibinfo
  {author} {\bibfnamefont {D.~T.~C.}\ \bibnamefont {Allcock}},\ }\href
  {\doibase 10.1103/PhysRevA.107.032413} {\bibfield  {journal} {\bibinfo
  {journal} {Phys. Rev. A}\ }\textbf {\bibinfo {volume} {107}},\ \bibinfo
  {pages} {032413} (\bibinfo {year} {2023})}\BibitemShut {NoStop}%
\bibitem [{\citenamefont {Boguslawski}\ \emph {et~al.}(2023)\citenamefont
  {Boguslawski}, \citenamefont {Wall}, \citenamefont {Vizvary}, \citenamefont
  {Moore}, \citenamefont {Bareian}, \citenamefont {Allcock}, \citenamefont
  {Wineland}, \citenamefont {Hudson},\ and\ \citenamefont
  {Campbell}}]{Boguslawski2023}%
  \BibitemOpen
  \bibfield  {author} {\bibinfo {author} {\bibfnamefont {M.~J.}\ \bibnamefont
  {Boguslawski}}, \bibinfo {author} {\bibfnamefont {Z.~J.}\ \bibnamefont
  {Wall}}, \bibinfo {author} {\bibfnamefont {S.~R.}\ \bibnamefont {Vizvary}},
  \bibinfo {author} {\bibfnamefont {I.~D.}\ \bibnamefont {Moore}}, \bibinfo
  {author} {\bibfnamefont {M.}~\bibnamefont {Bareian}}, \bibinfo {author}
  {\bibfnamefont {D.~T.~C.}\ \bibnamefont {Allcock}}, \bibinfo {author}
  {\bibfnamefont {D.~J.}\ \bibnamefont {Wineland}}, \bibinfo {author}
  {\bibfnamefont {E.~R.}\ \bibnamefont {Hudson}}, \ and\ \bibinfo {author}
  {\bibfnamefont {W.~C.}\ \bibnamefont {Campbell}},\ }\href {\doibase
  10.1103/PhysRevLett.131.063001} {\bibfield  {journal} {\bibinfo  {journal}
  {Phys. Rev. Lett.}\ }\textbf {\bibinfo {volume} {131}},\ \bibinfo {pages}
  {063001} (\bibinfo {year} {2023})}\BibitemShut {NoStop}%
\end{thebibliography}%

\clearpage

\onecolumngrid
\renewcommand{\thefigure}{S\arabic{figure}}
\setcounter{figure}{0}
\renewcommand{\theequation}{S.\arabic{equation}}
\setcounter{equation}{0}
\renewcommand{\thetable}{S.\Roman{table}}
\setcounter{table}{0}
\section{Supplemental Information}
\section{m-qubit Operations}
\begin{figure}[h]
    \centering
    \includegraphics[width=\textwidth]{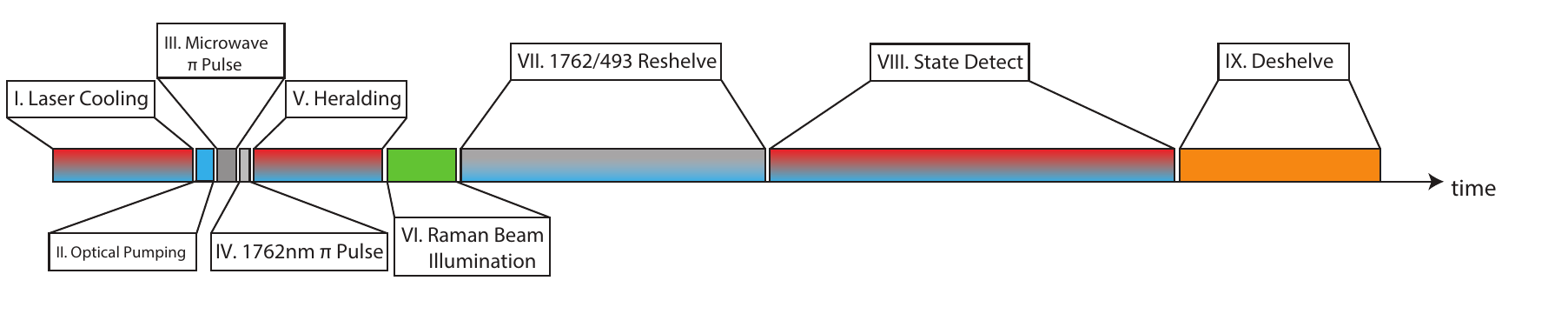}
    \caption{Experimental sequence diagram of metastable operations. }
    \label{fig:metaseq}
    \vspace{-10pt}
\end{figure}
\subsection{Heralding and D$_{3/2}$ Electron Shelving Sequence}
The narrow electric quadrupole 1762 nm optical transition is used to transfer population between the $g$ and $m$ qubit manifolds. The experimental sequence can be seen in Figure \ref{fig:metaseq}. \newline
\textbf{I.} Laser cooling for 8 ms with 493 and 650 nm light, both implementing an electro-optical modulator (EOM) to produce multi-tone light resonant with the hyperfine splitting of the S$_{1/2}$ and D$_{3/2}$.\newline
\textbf{II.} 493 nm-light hyperfine tones are turned off and a 1840 MHz tone resonant with the $|\mathrm{S}_{1/2},F=1,m_F=0\rangle$ to $|\mathrm{P}_{1/2},F=1\rangle$ populates the $|\mathrm{S}_{1/2},F=0,m_F=0\rangle$ state in 150  $\upmu$s. \newline
\textbf{III.} $130\,\upmu$s microwave $\pi$ pulse from $|\mathrm{S}_{1/2},F=0,m_F=0\rangle$ to $|\mathrm{S}_{1/2},F=1,m_F=0\rangle$. \newline
\textbf{IV.} A 45 $\upmu$s 1762 nm pulse transfers around 90-95\% population to the $|\mathrm{D}_{5/2},F=3,m_F=0\rangle$ state. \newline
\textbf{V.} To check that population has been transferred to the $m$ qubit manifold, the cooling lasers are turned on for 8 ms and photon counts are collected. If phonon counts are below a discriminating threshold, the state has been transferred. If photon counts are above this threshold, the population has remained in the $g$ qubit manifold and that experimental cycle is thrown away. \newline
\textbf{VI.} Two, co-propagating, detuned, 532 nm beams illuminate the ion, leading to coherent transfer of population between $|\mathrm{D}_{5/2},F=3\rangle$ and $|\mathrm{D}_{5/2},F=2\rangle$ hyperfine clock states. The frequency of a single beam is varied with an acousto-optical modulator (AOM) to produce data seen in Fig \ref{fig:metaseq_scan}. \newline
\textbf{VII.} To read out the $m$ qubit state we selectively shelve the $|\mathrm{D}_{5/2},F=3\rangle$ hyperfine clock state by turning on the 1726 nm laser along with the 493 nm beam for 7 ms. The combination of these lasers shelves this state to the D$_{3/2}$ manifold with a fidelity of 0.9989(2). \newline
\textbf{VIII.} The cooling lasers are turned on for 10 ms to measure photon counts. All population that was originally in the $|\mathrm{D}_{5/2},F=3\rangle$ will have been shelved to the D$_{3/2}$ manifold via step VII, therefore projecting these measured states as bright (high photon counts). Any population not affected by step VII would have been in the $|\mathrm{D}_{5/2},F=2\rangle$ hyperfine clock state and will be dark (low photon counts). \newline
\textbf{IX.} A 614 nm laser resonant with the D$_{5/2}\longrightarrow \text{P}_{3/2}$ manifold returns any population in the $m$ qubit subspace back to the $g$ qubit subspace to be Doppler cooled at the start of the next experimental cycle. 
\subsection{Measurement of $\mathrm{D}_{5/2}$ Hyperfine clock state Splitting}
\begin{figure}[]
    \centering
    \includegraphics[width=.8\textwidth]{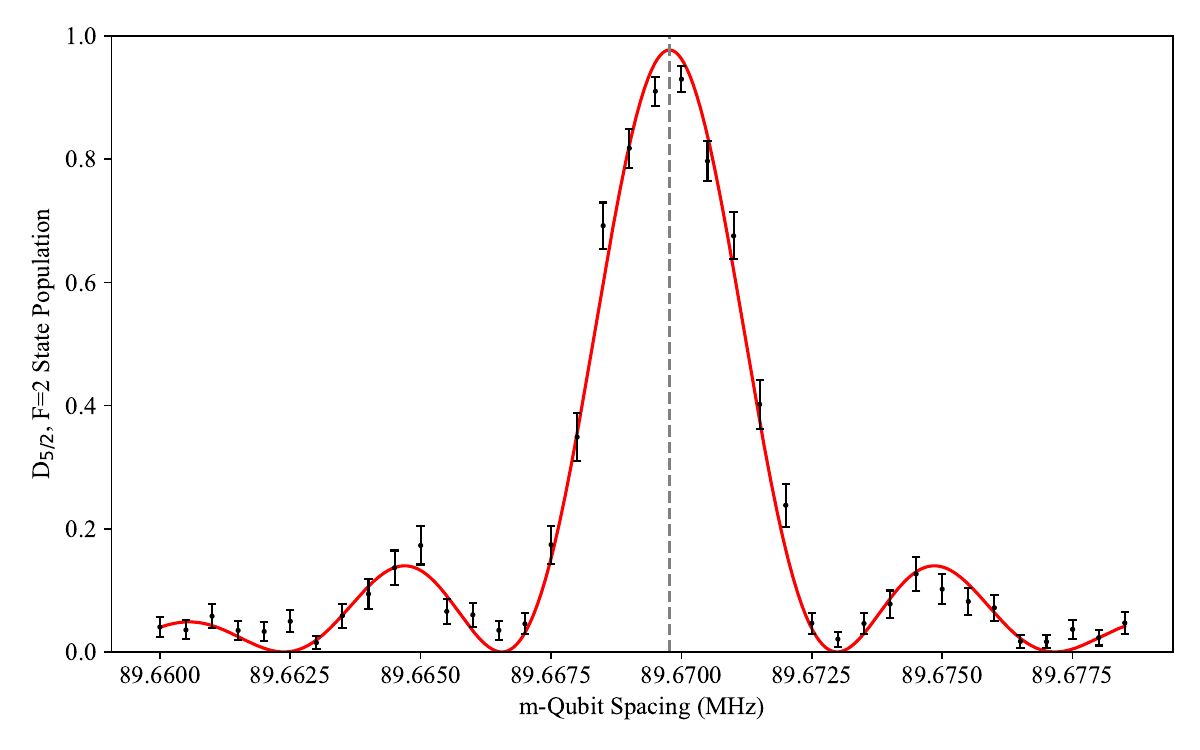}
    \caption{Experimental measurement of the $\mathrm{D}_{5/2}$ hyperfine splitting using high power, 532 nm laser to drive a stimulated Raman transition}
    \label{fig:metaseq_scan}
    \vspace{-10pt}
\end{figure}

Figure \ref{fig:metaseq_scan} shows a frequency scan of the two beam splitting of the Raman gate beams in step VI in the experimental sequence (see figure \ref{fig:metaseq}). This frequency splitting is realized by two AOMs, where one is double passed such that changing the driving frequency does not change the Poynting vector of the beam at the ion. Applying 125 MHz to the double pass creates a +250 MHz tone on the first order of the output of the double pass. Spatially overlapping, this beam with the positive-frequency offset diffraction from a second AOM driven at 160 MHz gives two tones with adjustable separation centered around 90 MHz. Running the sequence described above and scanning the double pass AOM frequency while keeping the illumination pulse time constant creates the spectroscopy data seen in figure \ref{fig:metaseq_scan}. The $\pi$-pulse time is 234 $\upmu$s and can be shortened with increased intensity. We choose to use a large  beam to increase the efficiency of the time needed to overlap both beams onto the ion.   The splitting measured here is 89.6697(4) MHz at 5.036(1) gauss. Accounting for the Zeeman shift of the D$_{5/2}$ levels, the zero field hyperfine splitting is approximately 89.2695(4) MHz. 
\section{g qubit Coherence Measurement}

\subsection{Detuned Ramsey Sequence}
To preform a differential shift and coherence measurement of the $g$ qubit we employ a detuned Ramsey sequence as follows: prepare the 0 state of the $g$ qubit, preform a detuned $\pi/2$ microwave pulse in the $g$ qubit, wait some period of time, and then preform another detuned $\pi/2$ microwave pulse. As the wait time is varied, a Ramsey fringe forms which will vary with frequency $f_\mathrm{Ramsey}$ which corresponds to the detuning of the microwave pulse from the true $g$ qubit hyperfine splitting. The sequence is repeated 300 times for each unique wait time. The decay of this fringe amplitude is fit to a decaying exponential give a coherence time of our $g$ qubit. 

As seen in figure 2 of the main text, we preform 3 different Ramsey fringe measurements. The top fringe is taken with no laser light on during the wait time, leading to no differential light shift of the $g$ qubit spacing. The frequency fit corresponds to a microwave shift of $440.4 \pm 0.13$ Hz. The fit gives a decoherence rate consistent with zero within the 30ms of measurement time.  

The middle panel is a sequence in which we illuminate the ion during the wait period with high power 532 nm light with a polarization of $A=-1$ ($\sigma^-$). The frequency of this Ramsey fringe is $f_\mathrm{Ramsey}=157.3 \pm 1.1$ Hz. The coherence time of the $g$ qubit illuminated by this light is measured to be $30.28 \pm 6.1$ ms. There are both short time scale and long time scale intensity fluctuations present during the illuminating of the ion, in other words, there is a distribution of fluctuations spanning from low to high frequencies. Fluctuations with timescales comparable to or shorter than the Ramsey interrogation time contribute strongly to decoherence, leading to a much shorter coherence time.  Low frequency fluctuations, likely due to thermal effects can be seen as slow drifts of roughly 20 Hz/60 min = 0.33 mHz/min. This represents a change in laser intensity of $1\cdot 10^7$~W/m$^2$ over a typical 60~minute Ramsey experimental cycle.

The bottom panel is taken directly after the data seen in the middle panel but with a ``magic'' polarization value set by rotating a quarter waveplate in an electronic Thorlabs rotation mount(PN. ELL14K). This sets the polarization to $A=0.212$ corresponding to elliptical polarization. Because this data is taken directly after the data seen in the middle panel, the laser intensity on the ion is roughly the same. Due to the tuning of this polarization to a ``magic condition'' we can minimize the differential light shift in the $g$ qubit and shield this manifold from high frequency intensity fluctuations. The $f_\mathrm{Ramsey}$ fit value for this fringe is 455.67 $\pm$ .1 Hz. This represents a roughly 15 Hz shift from the no light on fringe seen in figure 2(a) of the main text. Due to experimental uncertainties regarding the magnetic field we are unable to fully cancel the differential shift. But even with this slight shift we are able to increase the coherence time from 30ms as seen in the middle panel to 3.5 $\pm$ 2.6 s for this ``magic condition'' curve.  The relatively large error of this time is due to the minimal decoherence that occurs during the allowed data taking time period for our experiment. 
To further quantize this effect we can calculate the differential light shift sensitivity of the $g$ qubit as 19 Hz/(kW/cm$^2$). Tuning the polarization of our $m$ qubit beams we are able to virtually negate the equivalent of 14.1 kW/cm$^2$ of laser intensity input onto the ion.

\subsection{Fitting Ramsey Curves}
To fit the Ramsey fringe curves we use a decaying exponential sin$^2$ function. The fitting equation is of the form:  
\begin{equation}
    A \cdot \exp\left(\frac{-t}{\tau}\right)*\sin^2(\pi\cdot f_\mathrm{Ramsey}\cdot t)
\end{equation}
where A is the starting amplitude of the fringe, $\tau$ is the coherence time and $f_\mathrm{Ramsey}$ is the frequency of the fringe. 

\subsection{Estimation of Beam Divergence}

In order to accurately determine the uncertainty in the angle between the lasers wave vector $\vec{k}$ and the measured magnetic field $\vec{B}$ we can calculate the divergence of the laser. The relevant equation is
$$\theta = \tan^{-1}\left(\frac{w-w_0}{2f}\right)$$
Where $w_0$ is the beam waist at the focal length, $w$ is the beam waist at the focusing lens and $f$ is the focal length. The beam is focused through a 400 mm focal length lens and the lens is positioned with a high accuracy 3D micrometer stage. Using this equation, along with the uncertainty associated with the coil current, we can estimate the divergence to be 13.5~mrad 0.8~degrees. 

\end{document}